# The Reification of an Incorrect and Inappropriate Spreadsheet Model

Grenville J. Croll
EuSpRIG – European Spreadsheet Risks Interest Group
grenvillecroll@gmail.com
v1.60

*Once information is loaded into a spreadsheet, it acquires properties that it may not deserve. These properties include believability, correctness, appropriateness, concreteness, integrity, tangibility, objectivity and authority. The information becomes reified. We describe a case study through which we were able to observe at close hand the reification of a demonstrably incorrect and inappropriate spreadsheet model within a small non profit organisation.*

## 1 INTRODUCTION

[Taleb, 2007] prosaically introduced and summarised the issue which informs the existence and content of this paper:

> *"... things changed with the intrusion of the spreadsheet. When you put an Excel spreadsheet into computer-literate hands you get a 'sales projection' effortlessly extending ad infinitum! Once on paper or on a computer screen .....the projection takes on a life of its own, losing its vagueness and abstraction and becoming what philosophers call reified, invested with concreteness; it takes on a new life as a tangible object"*

We first and peripherally encountered the reification of spreadsheet information following the financial crash of 2008 [Croll, 2009]:

> *"the financial valuations expressed in CDO spreadsheet models were reified in the manner summarised earlier by Taleb".*

The discussion [Myers, 2005] of the study [Thorne, 2012] of the spreadsheet related problems in the UK nuclear fuel industry notes:

> *"At the centre of the incident are the computer spreadsheets, reified forms of the process of measurement…that was also what enabled the process to be subverted"*

There is otherwise a total absence of documented experience of the phenomena of reification of spreadsheet information within the literature. The purpose of this paper is to document a relevant case study which recently became available via a small non-profit organisation in the hope that it assists others in their documentation, investigation and management of the phenomenon.

In this paper we introduce the organisation and its historical finances, the financial planning spreadsheet that is the focus of this study, the spreadsheet error issues that were raised and the decision processes they influenced. We provide a summary which includes some learning points which may benefit others in a similar situation. We have anonymised the organisation and its domain of operation, officers and members.

## 2 LEISURE SERVICES ORGANISATION

### 2.1 Introduction

This paper is based upon a small non-profit leisure services organisation (LSO). The organisation is a UK limited company and thus its financial accounts are publicly available. It has Community Amateur Sports Club (CASC) status from which it derives some public financial privileges, in particular freedom from taxation on profits. It has been established over 40 years and has been historically grant aided by the UK Sports Council and the UK National lottery. The LSO has historically received other gifts, bequests and donations from a variety of sources. The LSO owns its own property from which it operates (thus it pays no rent) and significant technical assets related to its domain of operation. It provides access to and use of these assets for the benefit of its members (of which there are 50-100) in exchange for annual membership and point of use fees. The domain of operation of the LSO is not relevant and the intention is that the organisation and its directors, officers and members remain anonymous. We have successfully used the cloak of anonymity in previous work [Croll, 2005]. The LSO is managed by a committee and controlled by a smaller number of directors including a chairman and a treasurer who also sit on the committee. All the management are volunteers, are members of the LSO and of some years standing. The management retire by rotation after periods of between three and five years. There has been a plurality of directors, officers, members and reporting accountants and auditors over the years. The management is transparent, with regular meetings and the prompt circulation of detailed minutes, decisions, financial, technical and operational data.

### 2.2 Key Financial Data

The LSO has been profitable since incorporation and has built up a significant cash and asset base to support its operations. Though initially indebted to a high street bank and one or two benign creditors, the LSO was grant aided in 1996 and received a significant cash bequest in 2001. These enabled the LSO to replace its by then dilapidated technical assets with more up to date assets for members enjoyment. The assets have performed well over the ensuing 20 year period to date (2016) with relatively minor maintenance requirements and have significant remaining life and financial value. The LSO was given another cash sum (a 50% contribution) within the last three years by an anonymous donor in order for it to replace a rented asset with a newer and more efficient asset of its own costing circa £60k plus a further £20k for restitution of the rented asset prior to its return.

### Table 1 – LSO Key Balance Sheet Data 1993 - 2016

|  | 1993 | 1996 | 2004 | 2016 |
|---|---|---|---|---|
| **FIXED ASSETS** | | | | |
| Tangible Assets | £156,000 | £212,000 | £252,000 | £247,000 |
| **CURRENT ASSETS** | | | | |
| Stock | | £0 | £0 | £3,000 |
| Debtors | £3,000 | £8,500 | £5,000 | £8,000 |
| Cash | £3,000 | £500 | £104,000 | £115,000 |
| **CREDITORS** | | | | |
| One Year | -£41,000 | -£20,000 | -£4,000 | -£10,000 |
| **NET CURRENT ASSETS/LIABILITIES** | -£35,000 | -£10,000 | £105,000 | £116,000 |
| **TOTAL ASSETS** | £121,000 | £201,000 | £357,000 | £364,000 |
| **CREDITORS** | | | | |
| More than one year | £45,000 | £0 | £0 | £0 |
| **CAPITAL AND RESERVES** | | | | |
| Income and Expenditure | £76,000 | £201,000 | £220,000 | £226,000 |
| Capital Reserve | | | £138,000 | £138,000 |
| **SHAREHOLDERS FUNDS*** | £76,000 | £201,000 | £358,000 | £364,000 |

We show in Table 1 key balance sheet data (rounded) for the key years of this analysis obtained from the public record [Companies House Beta, 2017] by 24 separate downloads. We show in Figure 1 the key data graphically and in more detail for the 24 year period 1993-2016. It is very likely that none of the members, directors or officers of the LSO have seen the LSO data as depicted in Table 1 and Figure 1, though parts of it will be necessarily familiar to some of them.

The balance sheet data shows the transition from early debt funding to grant aided "equity" funding, modest book profitability over the long term and stable cash, capital and reserves. Trade stocks, creditors and debtors are minimal due to the nature of the organisation. Note that the company is limited by guarantee and has no shareholders. The rest of this analysis focuses upon the years 1996-2016 following acquisition of the main technical assets.

**Figure 1 – LSO Key Financial Data**

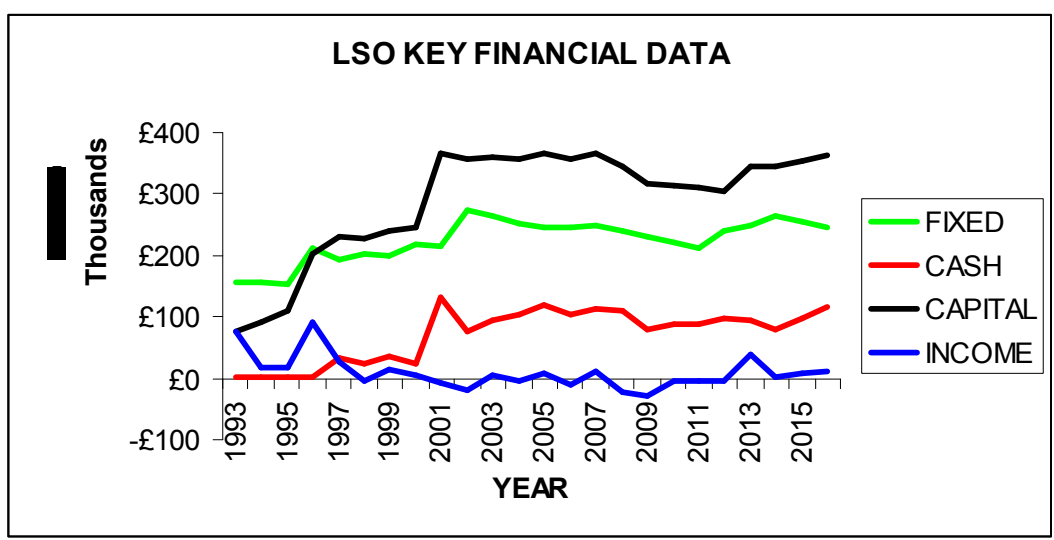

The technical assets, with the exception of the recent acquisition noted above, are fully depreciated with a Net Book Value of effectively zero. All the technical assets have calculated or estimated remaining lives of the order of 20-30 years. The technical assets have an estimated fair market value of £150k and are easily and regularly tradable in a pan-European marketplace. Thus the public accounts significantly understate the assets (and thereby profits) of the company, but are nevertheless properly prepared under the historical cost convention. The near term sale of the technical assets as part of an asset upgrade process would crystallise these profits or they could simply be revalued as happens on an annual basis at other LSO's of this type.

The LSO's freehold property was acquired in the late 1980's and is represented at £195k book value in the accounts and is not depreciated as a matter of policy. Some historical property improvements - £60k - have been capitalised and are included in this figure. The recent profits of the company are also understated because further amounts of property improvement have been booked to the P&L. These uncapitalised discretionary property improvements amount to a further £60k.

We show in Table 2 our summary of the LSO's estimated actual profits before depreciation 1996-2016.

**Table 2 – LSO Estimated Profits Before Depreciation**

|                          | 1996-2016 |
|--------------------------|-----------|
| INCOME AND EXPENDITURE   | £25,000   |
| DEPRECIATION             | £126,000  |
| RESIDUAL ASSET VALUE     | £150,000  |
| UNCAPITALISED PROPERTY   | £60,000   |
| TOTAL                    | £361,000  |
| PER ANNUM                | £18,050   |

A recent independent review of the LSO's financial statements by the Development Officer of the LSO's umbrella organisation confirms that the LSO is "*in a good financial position*". It is one of only a handful in the United Kingdom to own its own property and thus pay no rent. Comparison of its financial statements with other LSO's of the same type and size would reveal that it is one of the wealthiest smaller LSO's of its type in the United Kingdom.

**4 THE FINANCIAL MODEL**

**4.1 Background**

During the last few years members expressed through a series of informal meetings with the directors and a series of resolutions at Annual General Meetings an increasing desire that the directors replace or upgrade the technical assets with more up to date technology. This became imperative as the number of technical assets available for members use decreased recently due to accidental damage, disposal or periods of more extended maintenance. The chairman, directors and management of the LSO were resistant to change and procrastinated at every opportunity. Eventually, the chairman decided to commission a financial review of the LSO which involved the production of an Excel based financial model by two of the committee members, one of whom was a director.

**4.1 Structure of the Financial Model**

The model was a simple annual cash flow covering the 35 year period from 2016 to 2050 inclusive. The major maintenance and eventual replacement costs for each of the technical assets were tabulated for each year 2016-2050 in 2016 Money Values using a simple mechanism to account for perceived periodicity in major maintenance activities. Thus for the first asset, we have "[refurbish]" estimated at £18k in 4 years followed by £18k every 30 years. So the model would switch in an expenditure of £18k in 2020 and £18k in 2050. Since the first asset has a finite life the second entry was its replacement cost of £50k in 24 years followed (apparently) by its re-replacement of £50k in a further 24 years (i.e. beyond the scope of the model). All the assets ran for their full estimated or calculated lives and were replaced with an asset worth the same as the present (2016) value of the asset. End of life asset values were assumed to be zero. All the other routine operating costs of the technical assets were ignored as these would be covered by point of use revenue which was not included in the model. Major improvement expenditures for the property assets were subject to the same treatment. There were some minor one off expenditures also included in the model and a binary variable for calculating the effect of adding or not adding in the costs of a much desired further technical asset with a net budgeted cost of £40k.

Against cash expenditure as outlined above, the model had the then estimated 2016 opening bank balance of £110k and three choices of future cash income: £4k per annum, £8k per annum and £12k per annum. These were three choices of "profit before depreciation and one offs" based upon the modellers' view of the company's historical trading and how this might continue into the future and provide an inbound cash flow. There was no narrative describing how these figures had been arrived at. The formulae within the model show that the figures £12k and £4k were +150% and -50% computations around the central figure of £8k, which figure is at variance with our estimate of £18k pa in Table 2.

The opening bank balance was credited and debited with income and expenditure for the first and then each subsequent year. The carried forward balances were created for each year until 2050 in 2016MV's. A separate row on the spreadsheet inflated the 2016 carried forward balances at an inflation rate given in the model at 2%. A safety bank balance initially set at £30k was also inflated through the model's period for comparative purposes.

**4.2 LSO Financial Model Output**

We show in Figure 2 & 3 the charts depicting the models output which were first presented and discussed at an LSO board meeting in October 2016. Figure 2 shows the model output assuming no further purchases of technical assets beyond end of life replacement and Figure 3 shows model output with an immediate purchase of a new technical asset.

**Figure 2 – LSO model output – no new technical asset**

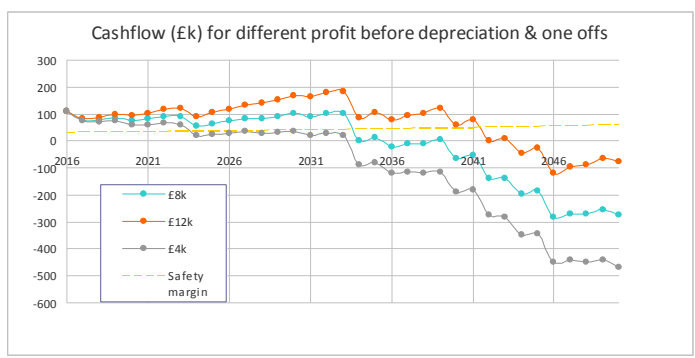

**Figure 3 – LSO model output – immediate purchase of new technical asset**

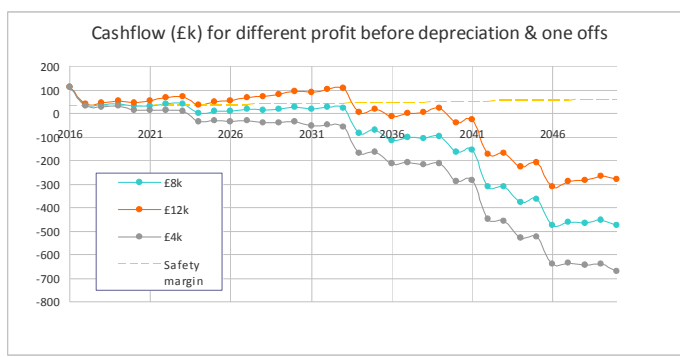

Figures 2 and 3 suggest an organisation in financial difficulties with medium term insolvency under all three income assumptions whether or not an additional technical asset was purchased. Figures 2 and 3 bear no relation to the history of the organisation obtained from the public record and summarised in Table 1 and Figure 1.

**4.3 Results of the LSO Financial Review**

A few days after the presentation of the financial model to the LSO board, the results of the financial review were communicated by the chairman to the members of the LSO in late October 2016:

> *"You will probably be aware that we have been carrying out a detailed review of … finances recently. The prime reason for this was to confirm what funds …. must reserve for necessary maintenance and, in due course, replacement of …. assets and so allow identification of what amount of the ….. current bank funds may be used to purchase an additional [technical asset]. …..A further reason for the review was to set up a system which will allow the [LSO] to monitor the financial condition of the [LSO] more easily. The findings of that review were presented to the Committee ….. so that we could consider the conclusions to be drawn from it.*
>
> *The [LSO] is fortunate to own a number of major [technical] assets. Most will have to be replaced at some point in the future; all will need major refurbishments and/or component replacement during their remaining working lives. We have spoken directly to suppliers… [etc, and]… have been able to arrive at a best judgement of when the major expenditure events will occur and what their costs are likely to be.*
>
> *…..This exercise has shown us that we will have to live within our means if we are to hand over to successive generations …. a financially healthy [LSO]. It is the unanimous conclusion of the Committee that we should not, at this juncture, purchase another [technical asset]. We believe that, unless or until, we can significantly improve our profitability, we should maintain the [technical assets as they now are]…..*
>
> *We intend to further refine the mechanisms used for the review and to continue to use them to update and monitor our financial status. The objectivity they provide will improve the ability of the Committee and Directors to take informed decisions in the best interest of the [LSO].*
>
> *If you would [like] more information on the financial modelling ……"*

Thus the information in the LSO's spreadsheet based financial model became reified. It was invested with confidence, correctness, authority, integrity, believability, appropriateness and even objectivity. It was set to become the mechanism by which the LSO was to be managed for the future. Most importantly, the review denied any material replacement or upgrading of the LSO technical assets despite a healthy bank balance (unused for two decades), historically secure finances and the members reasonable wishes formally and informally expressed over a prior period of several years.

**4.4 Initial challenge to Financial Model**

In the initial absence of the financial model itself and the data presented in Table 1, a challenge to the conclusions derived from the financial model was made based upon the information published in the 2015 statutory accounts which were the latest available at the time:

*"....we have conducted a detailed examination of the [LSO]'s financial situation by downloading the [2015] statutory accounts of [LSO] from Companies House.....*

*These accounts show that the current assets of the [LSO] - principally cash - have increased this last accounting year by £16k from £80k to £96k. This is largely due to the profitable activity of the [LSO] as reflected by the increase in its income and expenditure account. The Net Assets of the [LSO] are £352k....*

*.....The [LSO]has no debt and is almost free of taxation due to its [CASC] status…*

*..In summary the statutory accounts of [LSO] show that the [LSO]is profitable, free of debt, has no taxation ... and has reasonably provided for the preservation of its assets through an appropriate charge for depreciation. It has net assets in excess of £350k and cash assets in excess of £95k as of its last statutory accounts. These assets have significantly increased since 2012....*

*[we are] curious about the spreadsheet based financial modelling which has taken place to demonstrate that we cannot afford further [technical assets] that might better match our needs or aspirations……*

The two modellers requested a meeting the following evening to discuss the financial model and the above review. Although recollection of the meeting is not exact, it is certain that the quality of the model and the data upon which it relied were positively asserted with undue confidence [Panko, 2003]. No mention was made of any testing [Pryor, 2004; Panko, 2000 & 2006] that had taken place and no documentation [Pryor, 2006] was produced. A copy of the financial model and the presentation made to the board was however promptly provided the next day.

**4.5 LSO Financial Model Review**

A review of the LSO Financial Model occurred in two phases, much as described in our earlier work [Croll, 2003]. The low level review was kindly performed by a third party not involved with the LSO.

**4.5.1 Low level review**

The low level review of the LSO financial model revealed no major errors, save for the observation that about half of the model was not in use. There were many examples of poor practice such as embedded constants, complex formulae, strange formatting and a failure to colour code inputs and outputs. There was no evidence of any testing or documentation save a small revision history and a few cell notes.

**4.5.2 Initial High Level Review**

An initial high level review of the model was rapidly performed and sent directly to the modellers and relevant officers of the LSO. Despite the relatively small size of the model there were five main areas of concern:

- The use of an exponentially inflated cash flow over an extended future period of 35 years for a small organisation which gave the misleading impression of the evisceration of cash. This point was badly worded in the document sent to the modellers.

- The use of an understated inflation rate of 2% instead of average inflation over the last 20 years, which is nearer to 3% [Bank of England, 2017] which would have "*highlight[ed] the absurdity of the model*"

- Failure to use the established methodology of Discounted Cash Flow whereby future cash flows are discounted back to the present using an appropriate discount rate in order to facilitate the comparison of investment alternatives. A brief explanation of the DCF methodology was included.

- Use of data derived from overly conservative depreciation rates which greatly understated the gross margin available for the replacement of capital assets.

- No allowance for management's ability to postpone, reduce, avert or obviate key capital actions through the use of common sense, technology or asset replacement.

The high level review pointed out that the model reified or made concrete potential actions that are in the distant future.

### 4.5.3 Model test script

Following investigation of the model at high and low level, a very simple test script was devised to demonstrate the models failings. The test script contained the exact keystroke by keystroke detail required to perform the following three tests:

- Use the model to calculate how much cash is required in 2016 in order for the LSO to meet the specified cash safety margin in 2050. This was £442k – four times the 2015 cash balance.

- Show how by changing the inflation rate assumption from 2% to 3% that the amount of cash required in 2016 to meet the 2050 safety margin increases to £639k – six times the 2015 cash balance.

- Show how, by changing the refurbish period for asset one from 30 years (occurs in 2050) to 31 years (would occur in 2051 but outside scope of model) reduces the cash required in 2016 by 35k, demonstrating "edge effects" in the model.

The test script also asked users to:

- Note that the model does not include or provide for cash (and profit) that can be generated by the sale of current fully depreciated assets. i.e. the estimated £150k value of the existing assets is not included in the model.

The test script was sent to the Chairman and Treasurer in January 2017 with a request that the model not be used as a basis for further decision making unless and until the tests had been performed. Unfortunately the model had already delivered its results via the Chairman to the members two months previously as described in section 4.3.

**4.6 Model re-implementation**

In order to confirm that the LSO financial model was free of material formula error, we re-implemented it using the original maintenance data provided. We were able to cut out the half of the model that was not used, eliminate the complex formulae used to switch in the periodic expenses and perform simple tests around discount rates and asset replacement strategies.

**4.7 Further review of the model**

During the course of the preparation of this paper the model was subjected to further examination during which the following additional observations were made:

- The model did not relate the easily available historical cash flow depicted in Table 1 and Figure 1 to the projected future cash flows of Figures 2 and 3.

- The model implemented a single "use till death of asset" strategy only. There was no investigation of the value or otherwise of immediately upgrading technical assets by selling them for cash and a book profit and then replacing them with newer and upgraded technology using the otherwise unused cash available. There are many ways of interpreting reality and implementing a spreadsheet model [Banks & Monday, 2002].

**4.8 Response to the High level reviews & the Test Script**

The chairman responded:

*".....Your fundamental objection to the failure of the model to use the method of discounted cash flow is correct..... I have used DCF for evaluation of projects and the interest rates that you mention are not distant from those that I have come across being proposed by funding agencies. However… to this extent the model is flawed but it does allow attention to be focussed on how best to operate financially and to give some indication of whether corrective action is required…. A sensitivity analysis is usually carried out to show robustness of predictions and this has not been done"*

The principal modeller responded:

*The model uses a central forecast of £8k for profit from core operations (before …. one-offs and depreciation). The draft 2015-16 financial accounts show a higher number. What's the right long term number? The model is sensitive to this number.*

And then:

*"Re cash flow, lets ignore the discount/inflation rate for the moment and just focus on the cash flows. The [Asset one] £50k in 2040 should be the net cash outflow of buying a replacement after selling the near end of life [Asset one], otherwise it can*

*be reduced by whatever we think the market price of [Asset one] will be after 25 yrs further use".*

The LSO director involved in the modelling responded to an explanation of Table 2 as follows:

*"We have tried to quantify what we can reasonably see coming down the track and to fund it we need cash, rather than theoretical profit or liquidating current assets, to buy things. We've used inflating costs and inflating income and, thereby, inflating profit. We've taken informed guesses at the relevant values which we intend to modify as better data becomes available"*

To which the reply:

*"....when the [LSO] decides to sell [assets] to get something else it will be a statutory requirement to book the sale value as a profit because the ..... assets are fully depreciated. This is not a theoretical profit. It is an actual profit. To financially report otherwise would be unlawful.*

*"...Assets are routinely sold to support the purchase of other assets. All the [LSO's ]assets are capable of being sold at some time or other and where these amounts are significant - to the tune of £150k - it would be negligent not to include them as part of a competent financial model designed to support the management of the [LSO]"*

received no response. It is likely that the director is not aware of the basic tenets of Generally Accepted Accounting Principles (GAAP) or the theory and practice of DCF.

An *ex officio* member of the committee responded:

*"I think it's reasonable to conclude that our model is not accurate, and basing decisions on a flawed model would be foolish"*

There was no response from the chairman and the treasurer to the test script and no evidence that it had been executed.

## 5 ALTERNATIVE FINANCIAL MODELS

We briefly investigated two alternative financial models which could bring clarity to the decisions faced by the LSO.

### 5.1 Discounted Cash Flow

This was identified as the preferred capital appraisal evaluation methodology in the high level review. In this application DCF has its own problems, not least the selection of an appropriate discount rate [Croll, 2010]. Also, for positive discount rates, DCF will force significant capital expenditures into the far future supporting the "use till death" strategy of the LSO financial model.

The financial model described in this paper was constructed such that simply changing the inflation rate to a negative number turned the model into a DCF model. The modellers did not appear to realise this. The UK government has traditionally used a discount rate of 6% nominal on publicly funded infrastructure projects. Private firms would use a higher

Proceedings of the EuSpRIG 2017 Conference "Spreadsheet Risk Management" ISBN : 978-1-905404-54-4
Copyright © 2017, EuSpRIG European Spreadsheet Risks Interest Group (www.eusprig.org) & the Author(s)

rate on the same projects. We show below the output of the LSO model with a 6% discount rate (entered as a -6% inflation rate) and a revised £18k inbound annual free cash flow per Table 2.

**Figure 4 – LSO model output – immediate purchase of new technical asset using 6% discount rate with revised £18k pa inbound future cash flow**

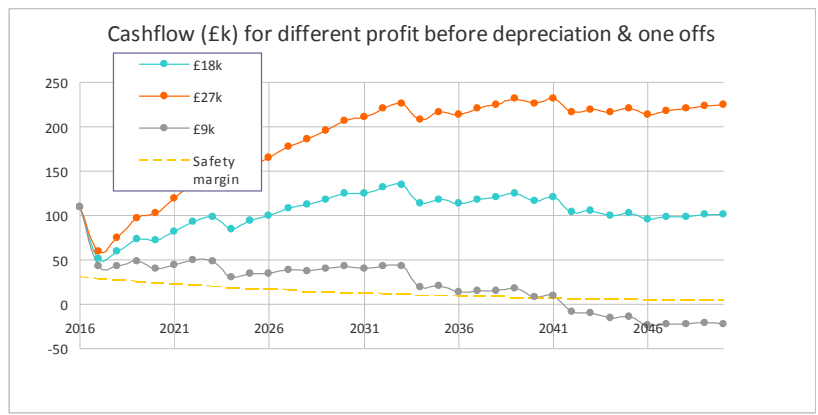

The DCF methodology using the same financial model, identical maintenance data, 6% nominal discount rate and a revised £18k annual inbound cash flow shows the organisation to be in good prospective health for the future. The projected future cash balances are similar to the existing and included the immediate purchase of a new technical asset, plus all the maintenance and improvements in the Financial Model.

**5.2 Return on Capital Employed (ROCE)**

The LSO Shareholders Funds for 1996 in Table 1 shows a balance of £201k. This increased to £364k in 2016. We might reasonably add to this 100% of the uncapitalised property improvement expenditure of Table 2 (£60k) and the fair market value of the fully depreciated technical assets of Table 2 (£150k). This suggests an adjusted Shareholders Funds account of £574k as at 2016. Thus the LSO achieved an adjusted ROCE of £574k / £201k = 285% over the 20 year period from 1996 or 5.4% compound per annum nominal, 2.8% real.

The LSO overall ROCE of +2.8% real contrasts markedly with the -2.8% per annum real that the LSO has achieved for its £133k cash balance over the period since 2001. The cumulative deflationary loss of capital that the LSO has suffered over the last 15 years due to its large cash holding of benefactors funds is approximately £69k in 2016 MV's, less a small amount of interest.

The ROCE methodology suggests that the LSO should immediately employ most of its available cash as technical assets in its own business save any cash contingency requirements.

The ROCE of 5.4% nominal supports a discount rate for use in the financial modelling of 6% nominal.

# 6 CONCLUSION

We have described a simple spreadsheet that was used in a financial review of a small Leisure Services Organisation [LSO] in order to evaluate the wisdom or otherwise of the upgrade, replacement or enhancement of its technical assets.

The spreadsheet presented to the management of the organisation suggested that the organisation was about to enter serious financial difficulties and it could not afford to upgrade or extend its assets. These projected circumstances were at variance with the solid historical financial performance of the organisation over the previous 20 years.

We have shown how the model became reified. That is to say the spreadsheet was invested with a variety of attributes including believability, correctness, appropriateness, concreteness, integrity, tangibility, objectivity and authority. None of these attributes were deserved as we were able to subsequently identify and describe numerous significant weaknesses and strategic flaws in the model. We revealed these issues through a straightforward software engineering process involving the performance of a low level model review, a high level model review, the production of a simple test script and the re-engineering of the model.

We investigated the historical circumstances of the LSO by downloading its publicly available accounts and transaction history for the prior 20 years and re-estimated its actual historical profitability. We show that the LSO's future projected free cash flow had been underestimated by a factor of more than two.

We then used the original model with a positive discount rate of 6% (rather than a negative discount rate of 2%) and our revised future free cash flow to perform a DCF analysis which showed that the LSO was easily capable of affording the new technical and other assets that its members aspired to.

It appears to be the case that the organisation had been planning its future over a 35 year period using a spreadsheet based Discounted Cash Flow Model with negative discount rates (contrary to centuries of tradition). In addition the model underestimated a critical spreadsheet variable by a factor of more than two.

We used a ROCE analysis to show that the LSO had generated a 5.4% nominal compound rate of return on its capital over the prior 20 years. This implied that the significant cash that it has had languishing in the bank would be far better employed for the future in the assets and business of the LSO.

The reification of the deeply flawed spreadsheet that we have examined has had the unnecessary and negative consequence of stagnating the organisation's future development. This lead one exasperated LSO member (a former LSO treasurer) to write:

> "The [LSO] and the wonderful facilities were built by adventurous forward looking people but the doom and gloom spreadsheet has permeated the soul of the [LSO] and has worked its insidious magic".

Our experience informs us that deeply flawed but reified spreadsheets have inflicted serious damage on other organisations too.

# 7 RECOMMENDATIONS

We significantly underestimated the amount of resource that would be required to challenge the simple spreadsheet model at the heart of this investigation. Those who follow, who might be working on spreadsheets a hundred times more complex, might wish to note the following:

- Prior to challenging an existing model, you will need to perform a full and effective high and low level model review and possibly develop a series of model test scripts. Any flaws in the review process or the test scripts will undermine the challenge.

- You may need to acquire or reacquire and then validate the relevant model source data.

- If the target model is strategically, tactically or computationally incorrect you may need to build another model in order to figure out what the right answer is in the first place. You will then need to perform a thorough high and low level model review on the challenger model. And then compare and contrast the differences between the challenger model and the existing model.

- Do not expect the management of the organisation you are dealing with to have the capacity or interest to comprehend the errors, limitations or flaws observed by you in their financial modelling or accounting methods.

Whilst the above processes are taking place – all at the same time, with great time urgency and in a seemingly haphazard event driven order - the original model (flawed in your opinion) at the heart of your challenge will be seeping deeper and deeper into the organisation.

Spreadsheets are ubiquitous and error prone but the scientific literature gives good guidance as to the processes required to investigate and correct the errors. It would be wise to follow the literature. We estimate that the time required to challenge a reified model is at least three to six times the time required to create the (probably untested) target model in the first place. This can be extraordinarily expensive and it may be cheaper and far less stressful to simply walk away and seek a fresh challenge elsewhere.

## ACKNOWLEDGEMENTS

We thank Patrick O'Beirne for performing the low level review at short notice. We thank an Auditor, a Software Engineer, a City of London Banker, a Fellow of the Irish Computer Society and a Professor of Law for their promptly provided informal reviews of this paper. We thank the anonymous referees for their helpful comments.